\documentstyle[12pt]{article}
\title{Neural networks with high order connections}
\author{Jeferson J. Arenzon and Rita M.C. de Almeida \\
Instituto de F\'{\i}sica \\
Universidade Federal do Rio Grande do Sul  \\ C.P. 15051 --
91501-970 -- Porto Alegre -- RS -- Brazil \\
E-mail: ARENZON@IF1.UFRGS.BR}

\begin{document}
\thispagestyle{empty}
\maketitle\vspace{-1cm}

\begin{abstract}

We present results for two different kinds of high order
connections between neurons acting as corrections to the
Hopfield model. Equilibrium properties are analyzed using the
replica mean-field theory and compared with numerical simulations.
An optimal learning algorithm for
fourth order connections is given that improves the storage
capacity without increasing the weight of the higher order
term. While the behavior of one of
the models qualitatively resembles the original Hopfield one, the
other presents
a new and very rich behavior: depending on the strength of the
fourth order connections and the temperature, the system
presents two distinct retrieval regions separated by a gap, as well
as
several phase transitions. Also, the spin glass states seems to
disappear above a certain value of the load parameter $\alpha$,
$\alpha_g$.
\end{abstract}

\newpage

\section{INTRODUCTION}

Synapses connecting more than two neurons have been introduced in
attempt to both improve the storage capacity
of existing models \cite{per}-\cite{trs} and to be a
simulacrum of synapses existing in real brains  (see
\cite{per} and references therein).
Biologically, the idea of multisynapses has a strong
motivation \cite{per}: axon-axon-dendrite
connections, for instance, are relatively common in real nervous
systems and can be described as third order synapses and even more
intricate  connections, involving more than two axons, may also
exist in the brain.  However, since second order synapses are
highly dominant, higher order terms should be considered as
corrections. As stressed in  ref.\cite{per}, this feature may play
an
essential role in the functioning of central nervous systems of
superior vertebrate organisms. Moreover, when some pairwise
connections are  close enough they may interact somehow, and  that
can also be considered as high order synapses (although in  this
case there are only interactions of even order).

Networks with $N$ infinite range interacting Ising spins ($S_i= \pm
1$)  associated to the state of the neurons (active or
inactive) are considered to describe  learning, storage, and
retrieval of information. Possible configurations  of the network
are represented by $N$-dimensional vectors
$\vec{S}=(S_1,\ldots,S_N)$ and the stored information (memories) is
associated to  $P$  of these states,  denoted by
the vectors $\vec{\xi}^{\mu}$, $\mu=1,\ldots,P$. The network load
is measured by the parameter $\alpha$, usually $P/N$,
and the performance of a model for attractor networks can be
measured by its storage capacity
and its ability in recalling the stored patterns, in particular,
the maximum allowed noise in an initial configuration and the time
needed by the network to evolve and stabilize at, or near, one of
the $P$ memories.

Several works introduced multispin interactions by generalizing the
Hopfield model \cite{hop} and Hebb learning rule by a monomial of
degree $k>2$ in the Ising spins \cite{multi,tama}. These models
have been investigated, both analytically  and
through computer simulations. Alternatively,
a recently introduced model \cite{rsana} simultaneously considers
several orders of interactions, besides the second order Hopfield
term. In this paper we present a  truncated version (hereafter
called  Truncated model) of that model, and study the effect of the
(weighted) first  correction to the Hopfield term. We also
investigate the effect of a  Hopfield-like correction (hereafter
called Generalized Hopfield model, GH) and compare both
prescriptions.

The paper is organized as follows: section 2 defines the models and
in section 3 analytical and numerical results are presented. In
section 4  we give an optimal learning rule that highly increases
the storage capacity  and finally in section 5 we summarize and
present our conclusions.

\section{THE MODELS}

\subsection{The Generalized Hopfield Model}

In order to compare performances, we contemplate a
straightforward generalization of the Hopfield model (GH)
by a polynomial of degree $M$
that considers multispin interactions \cite{multi,tama}, namely
\begin{equation}
E=-\frac{N}{2} \sum_{\ell=2}^M \varepsilon_{\ell} \sum_{\mu}
m_{\mu}^{\ell}  \:\:\:\:\: ,
\label{ghenerg}
\end{equation}
where $M$ is an integer ($M>2$) and the overlap $m_{\mu}$ between
the state of the
network $\vec{S}$ and the pattern $\vec{\xi}^{\mu}$ is given by
\begin{equation}
m_{\mu}=\frac{1}{N} \sum_{i=1}^N \xi_i^{\mu} S_i \:\:\:\:\:.
\end{equation}
Here we have high order
terms as  corrections (that do not need to be small)
to the original second order Hopfield model. This system has a
completely  different static and dynamical behavior from the one
that will be defined in the next subsection \cite{trs}. Here, the
higher order corrections do not qualitatively change the $T=0$
behavior of the Hopfield network although the $(\alpha,T)$ phase
diagram  presents some new  features. The Hopfield
network has its performance determined by the nature of the local
field on a given neuron $S_i$. This local field has two competing
components, a signal term that tends to align the spin $S_i$ with
a given pattern, and a noise term that has
a random orientation. A correction to the second order
Hopfield term may then act in two different ways: either it
enhances the signal term and/or it decreases the noise one. Also,
the joint analysis of both terms yields an estimate
of the critical capacity of the net: the lowest order
interaction ($\ell_{min}$) in the energy function  is the most
relevant  contribution for the cross talk noise from the high
patterns implying that the maximum number of patterns that can  be
embedded is ${\cal O}(N^{\ell_{min}-1})$. Thus, the presence of
the second
order term  implies that the maximum number of patterns that can be
stored is  proportional to $N$ \cite{trs,signal}.

The learning rule, for any $\ell\geq 2$, is:
\begin{equation}
J_{i_1\ldots i_\ell}^H (P+1) = J_{i_1\ldots i_\ell}^H (P) +
\frac{\varepsilon_\ell}{N^{\ell-1}}
\xi_{i_1}^{P+1} \xi_{i_2}^{P+1} \ldots \xi_{i_\ell}^{P+1}
\:\:\:\:\: . \label{klearning}
\end{equation}
Here one has the full symmetry of indices and all connections are
symmetric. Remark that in a more general case, the weights could be
considered different for each pattern,
$\varepsilon_{\ell}=\varepsilon_{\ell}(\mu)$, generalizing the
model studied by Viana \cite{laura}.

This model presents an overall behavior qualitatively similar to
the standard Hopfield model at $T=0$.
For $\alpha < \alpha_c$, the retrieval quality is good ($m
\simeq 1$) and the size of the basins of attraction decreases with
$\alpha$. Also, if the initial state is out of the basin of
attraction, the mean convergence time grows as the network
increases,  while inside the basins just one or two steps are
enough \cite{trs,ko}. Underlying this similarity is the fact that
the loss of retrieval abilities due to an overload of
the network originates in the same mechanism: the noise
term overwhelms the signal one when $\alpha>\alpha_c$.

\subsection{The Truncated Model}

The complete energy function of a model previously proposed
\cite{rsana},  with all orders of interactions (up to $2P$), for a
network   storing $P$ patterns is
\begin{equation}
E = N \prod_{\mu=1}^P \left( 1 - m_{\mu}^2 \right) \:\:\:\:\: .
\label{ers}
\end{equation}
This energy function is proportional to the product of  the Hamming
distances between the network state $\vec{S}$ and the patterns
$\vec{\xi}^{\mu}$ and its inverses $-\vec{\xi}^{\mu}$. From
eq.(\ref{ers}) it is clear that $E(\vec{S}) \geq 0$ (if
$\vec{S}=\vec{\xi}^{\mu}$, for any $\mu$, the equality holds). It
means that, no matter
how large $\alpha$ is, the patterns are always global minima of
$E$. A complete discussion of the phase space landscape in  the
$\alpha\to 0$ limit as well as  simulation results for  $\alpha\neq
0$ can be found in refs.\cite{rsana,rssim}.

The multi-interaction nature of equation (\ref{ers}) becomes
evident when it is displayed as
\begin{equation}
E=N \left[ 1 - \sum_{\mu_1} m_{\mu_1}^2 + \sum_{\mu_1<\mu_2}
m_{\mu_1}^2 m_{\mu_2}^2 + \ldots + (-1)^P \sum_{\mu_1 < \ldots <
\mu_P}  m_{\mu_1}^2 \ldots m_{\mu_P}^2 \right] \:\:\:\:\:.
\label{expan}
\end{equation}
Notice that although the first non-trivial term is
the Hopfield energy function, the higher order ones  are different
from any previous model because they contain mixed memory terms.
Also, since we have the Hopfield term along with higher order
ones, we expect that the number of patterns that can be stored is
${\cal  O}(N)$. We now
define an energy function by neglecting the constant zeroth
order one and considering the next $M$ terms ($M<P$). This energy
function, after introducing weights and renormalizing it by a
factor $1/2$, reads \begin{equation}
E=\frac{N}{2} \sum_{\ell=1}^M (-1)^{\ell} \varepsilon_{2\ell}
\sum_{\mu_1 < \ldots < \mu_{\ell}} m_{\mu_1}^2 m_{\mu_2}^2 \ldots
m_{\mu_\ell}^2 \:\:\:\:\:.
\label{tobetruncated}
\end{equation}
Equation (\ref{tobetruncated})  defines  a model that
can be  regarded as the Hopfield model ($\ell=1$) plus
correction terms. Differently from the previous case, here
one cannot have $\varepsilon_2=0$ because the remaining
terms are mixtures and no pattern can be retrieved only
with them. In what follows we consider only the first correction
to the Hopfield term ($M=2$). Eq.(\ref{tobetruncated}) can then be
rewritten as ($\varepsilon_{2\ell}=\delta_{1\ell} + \varepsilon
\delta_{2\ell}$)
\begin{equation}
E=-\frac{1}{2} \sum_{i,j} J_{ij} S_i S_j +  \frac{\varepsilon}{2}
\sum_{i,j,k,l} J^T_{ijkl} S_i S_j S_k S_l \;\;\;\; .
\label{liapunov}
\end{equation}
The learning rule for the second order couplings $J_{ij}$ is the
Hebb prescription \cite{model}, eq.(\ref{klearning}). The fourth
order synapses
$J^T_{ijkl}$, on the other hand, may be implemented
through the following learning rule
\begin{equation}
J_{ijkl}^T = \frac{1}{2N^3} \sum_{\mu\neq\nu} \xi_i^{\mu}
\xi_j^{\mu} \xi_k^{\nu} \xi_l^{\nu}
\end{equation}
and when a new pattern is learned,
\begin{equation}
J_{ijkl}^T(P+1) = J_{ijkl}^T(P) +
\frac{\varepsilon}{N^2} J_{ij}(P)
\xi_k^{P+1} \xi_{l}^{P+1} \:\:\:\:\: ,
\label{learning}
\end{equation}
where $\vec{\xi}^{P+1}$ is the $(P+1)$th pattern to be taught to
the net. Eq.(\ref{learning}) may be regarded as the multisynapses
described in the introduction: the last term being the action of
two axons upon a binary synapse. Notice that for $P=1$ this  model
is equivalent to the standard Hopfield one since the fourth order
synapses do not exist. Remark that while the pairwise
connections are symmetric, i.e. $J_{ij}=J_{ji}$, the fourth order
ones do not present full symmetry under all possible indices
interchange (only $i\leftrightarrow j$ and
$k \leftrightarrow l$). These couplings can be symmetrized if
they are rewritten as
\begin{equation}
J_{ijkl}' = \frac{1}{3} \left( J_{ijkl} + J_{ljki} + J_{kjil}
\right) \end{equation}
and hence, the energy (\ref{liapunov}) is a Lyapunov
function for the dynamics
\begin{equation}
S_i(t+1) = \mbox{sgn} \left( \sum_j J_{ij} S_j(t) - \varepsilon
\sum_{j,k,l} J_{ijkl}' S_j(t) S_k(t) S_l(t) \right)
\label{dynamic}
\end{equation}
and the use of the Statistical Mechanics tools is allowed.
One should also notice the importance of the self couplings here.
For instance, the couplings $J_{iikl}$ and $J_{ijkk}$ give rise to
contributions to the energy of the same order
as the couplings $J_{ij}$, what does not happen
in the generalized Hopfield model.
This can be easily seen if one rewrites the couplings as
\begin{equation}
J^T_{ijkl} = \frac{1}{2N} J_{ij} J_{kl} - \frac{1}{2}
J^H_{ijkl}\;\;\;\;.
\end{equation}
Then, for the self couplings mentioned above:
\begin{equation}
J^T_{iikl} = \frac{\alpha}{2N} J_{kl} - \frac{1}{2N^2} J_{kl}
\simeq \frac{\alpha}{2N} J_{kl} \;\;\;\;,
\end{equation}
where the factor $N^{-1}$ is compensated by a sum over
sites ($i$) in (\ref{liapunov}). Notice that the contribution from
$J^H_{iikl}$ is neglectable.

A previous numerical simulation \cite{trs}  for $\varepsilon=1$
shows a continuous transition from the retrieval phase to a
non-retrieval one at $T=0$. The basins of attraction seem  to be large
and $\alpha$-independent: $m_o^c$
(the minimum initial overlap that allows retrieval in the
thermodynamic limit) is $\sim 0.1$ for all $\alpha<\alpha_c$.  The
mean convergence time
$\langle\; T \;\rangle$ (the average number of whole network
updatings required to reach a stable state) increases with
$\alpha$ and, for non small values of $\alpha$,
does not depend on the initial overlap
$m_o$. Long convergence times and large dispersions  on its average
values are often related to the irregularity of phase
space  around  the memories
\cite{ko} (existence of spurious states). However, this
interpretation is only
valid when the stored patterns are at, or very near, the bottom of
the basins of attraction ($m \simeq 1$). Since the transition here
is continuous, for large values of $\alpha$ this is no longer true:
initial states with overlap $m_o$ with the chosen pattern lay on
the  surface of a hyper-sphere of radius proportional to $1-m_{o}$
centered at the pattern. On average, the distance in the phase
space from this surface to the bottom of the basin of
attraction is the same as the distance from the memory to the
energy minimum, independently of $m_o$. Consequently, the
convergence time does not depend on $m_o$. For small values of
$\alpha$, on the other hand, the retrieval quality is good ($m
\simeq 1$) and a small decrease in the convergence time is
observed as $m_o$ increases \cite{trs}.

\section{MEAN FIELD THEORY}

\subsection{The Free Energy and Saddle-Point Equations}

The mean field analysis is performed by means of the standard
techniques introduced by Amit {\it et al.}\cite{amit}.
In the GH model, the cross-talk noise is governed by the second
order term while the fourth order one contributes only to the
signal term. On the other hand, in the Truncated model there is a
contribution from the higher order terms (for instance, from the
self couplings
mentioned in the previous section) to the overall noise due to
microscopically overlapping patterns. Up to the fourth  order
($\varepsilon_{2\ell}=\delta_{1\ell} +\varepsilon
\delta_{2\ell}$), the Truncated energy function can be rewritten as
\begin{equation}
E = - \frac{N}{2}\sum_{\mu} m_{\mu}^2 -\frac{N\varepsilon}{4}
\sum_{\mu} m_{\mu}^4 + \frac{N\varepsilon}{4} \left( \sum_{\mu}
m_{\mu}^2 \right)^2 \;\;\;\;\;,
\label{et}
\end{equation}
and the free energy per neuron can be obtained using the replica
trick. After assuming that the
replicas are symmetric and taking the limit of zero replicas we get
\begin{eqnarray}
f_T = &-& \frac{1}{2} (1-\varepsilon y) \sum_{\mu} m_{\mu}^2 -
\frac{\varepsilon}{4} \sum_{\mu} m_{\mu}^4 - \frac{\varepsilon}{4}
y^2 + \sum_{\mu} t_{\mu} m_{\mu} \nonumber \\*
&+& \frac{\alpha}{2\beta} \ln [ 1-
\beta(1-\varepsilon y)(1-q)] - \frac{1}{2}
\frac{\alpha q  (1-\varepsilon y)}{1- \beta(1-\varepsilon y)(1-q)}
\nonumber \\* &+& \frac{1}{2}  \alpha\beta r (1-q)  -
\frac{1}{\beta} \: \left\langle \! \left\langle \: \ln 2 \cosh \beta
\left[  {\bf t}.\mbox{\boldmath $\xi$} + \sqrt{\alpha r} z \right] \:
\right\rangle \! \right\rangle \: \;\;\;\;,
\label{freeenergytrs}
\end{eqnarray}
where $y$ is introduced to linearize the last term in eq.(\ref{et})
and the variables  $t_{\mu}$, $q$, and $r$ are usually introduced
to linearize the non-linear terms. The symbol $\: \left\langle \!
\left\langle \:\;\;\: \right\rangle \! \right\rangle \:$
stands for two averages: over the finite number of patterns that
may condense and over the gaussian variable $z$, related to the
infinite number of microscopically  overlapping memories. The
saddle point equations are
\begin{eqnarray}
{\bf m} &=&  \left\langle \! \left\langle  \mbox{\boldmath $ \xi$}
\tanh \beta\left(
\mbox{\boldmath $ \xi$}.{\bf t} + \sqrt{\alpha r} z \right)
\right\rangle \! \right\rangle
\label{mtrs}
\\* t_{\mu}&=& (1-\varepsilon y)m_{\mu}+\varepsilon m_{\mu}^{3}
\\* q &=&  \left\langle \! \left\langle  \tanh^2 \beta
\left( {\bf t}.\mbox{\boldmath $\xi$}
+ \sqrt{\alpha r} z \right)  \right\rangle \! \right\rangle
\label{qtrs} \\* r &=& q\left[
\frac{1-\varepsilon y}{1-\beta(1-\varepsilon y)(1-q)}\right]^2
\\* y &=& \sum_{\mu}
m_{\mu}^2 + \alpha \frac{1-\beta(1-\varepsilon y)(1-q)^2}{\left[ 1-
\beta(1-\varepsilon y)(1-q) \right]^2}\;\;\;\;\;\; .
\label{sprs}
\end{eqnarray}

Analogously, for the GH model the free energy is
\begin{eqnarray}
f_H = &-& \frac{1}{2}
\sum_{\ell=2}^M \sum_{\mu} \varepsilon_{\ell} m_{\mu}^{\ell}  +
\sum_{\mu} t_{\mu} m_{\mu} + \frac{\alpha}{2\beta} \ln [ 1-\beta(1-
q)]   - \frac{1}{2}
\frac{\alpha q}{1- \beta(1-q)} \nonumber \\*
&+& \frac{1}{2}  \alpha\beta r (1-q)  -
\frac{1}{\beta} \: \left\langle \! \left\langle \: \ln 2 \cosh
\beta \left[  {\bf
t}.\mbox{\boldmath $\xi$} + \sqrt{\alpha r} z \right] \:
\right\rangle \! \right\rangle \:
\;\;\;\; ,\;\; \varepsilon_2 \neq 0 \;\;\;\; ,
\label{freegh}
\end{eqnarray}
and the equations to be solved are the
same as in the original Hopfield model \cite{amit} except for
$t_{\mu}$. The equations for {\bf m} and $q$ are the same as
eqs.(\ref{mtrs}) and (\ref{qtrs}) and $t_{\mu}$ and $r$ read (here
we  do not have $y$)
\begin{eqnarray}
t_{\mu}&=& \frac{1}{2} \sum_{\ell=2}^M  \varepsilon_{\ell} \ell
m_{\mu}^{\ell-1}  \label{tgh} \\*
r &=&  \frac{q}{\left[1-\beta(1-q)\right]^2} \;\;\; ,
\varepsilon_2 \neq 0 \;\;\;\; .
\label{mh}
\end{eqnarray}

The sets of coupled nonlinear equations given by eqs.
(\ref{mtrs})--(\ref{sprs}) and (\ref{mtrs}),(\ref{qtrs}),
(\ref{tgh})--(\ref{mh}), are
numerically solved in the next subsections in the case where the
network presents a macroscopic overlap $m$ with one of the
memories ($m_{\mu} = m \delta_{1\mu}$). Since we are mainly
interested in  the properties of the Truncated model, results for
the GH model will be given when they differ from the original
Hopfield model  or when comparing both models.

\subsection{The $T=0$ Limit}

The $T=0$ ($\beta\to\infty$) limit of the saddle point equations of
the Truncated model is
\begin{eqnarray}
m &=& \mbox{erf} \left( \frac{t}{\sqrt{2\alpha r}} \right)
\label{mt0} \\* t
&=& (1- \varepsilon y)m +\varepsilon m^3 \\*
 r &=& \left[ \frac{1-\varepsilon y}{1-C(1-\varepsilon y)}
\right]^2 \\*  y &=& m^2 + \frac{\alpha r}{(1-\varepsilon y)^2} \\*
C &=& \sqrt{\frac{2}{\alpha\pi r}} \exp \left( -
\frac{t^2}{2\alpha r} \right) \;\;\;\;,
\label{ct0}
\end{eqnarray}
where $q\to 1$ and $C\equiv \beta(1-q)$. These equations are
numerically solved for several values of $\alpha$ and $\varepsilon$
and the results are compared with the simulation whose details are
presented later. We found essentially two different regimes,
depending on the value of $\varepsilon$. For large $\varepsilon$
($\sim 0.5$) the  overlap $m$ decreases with $\alpha$, going
monotonically from 1 down to zero at $\alpha_c^+(\varepsilon)$,
signalling a second order phase transition.  As $\varepsilon$
decreases ($\sim$ 0.36),  the overlap presents a local minimum,
before finally going to zero at $\alpha_c^+ (\varepsilon)$. If
$\varepsilon<\varepsilon_c\simeq 0.3587$, the minimum yields a gap
separating two retrieval regions. These results are illustrated in
figs.\ref{trsm} and \ref{trsm2} and summarized in the $T=0$ phase
diagram of fig.\ref{pdt0}. The critical values
$\alpha_c^{\pm}(\varepsilon)$ are associated to second order
transitions at $T=0$.  The
gap is delimited by the lines $\alpha_c^-(\varepsilon)$ and
$\alpha_c'(\varepsilon)$, which meet at an endpoint
nearby (0.3543,0.7784) (see fig.\ref{cp}): the left border is
always given by $\alpha_c'(\varepsilon)$ while the right one is
defined by $\alpha_c^-(\varepsilon)$ (second order) for
$\varepsilon < 0.3543$  and by
$\alpha_c'(\varepsilon)$ (first order) for
$0.3543 < \varepsilon< 0.3587$. As $\varepsilon$ approaches zero,
the gap width, $\alpha_c^- - \alpha_c'$, goes to
infinity and for negative values of $\varepsilon$
only the first order transition associated to
$\alpha_c'(\varepsilon)$ is present (see fig.\ref{pdt0}). Near
$\varepsilon_c$ the behavior of the model is very complex near the
gap and  the effects of  replica symmetry breaking (RSB) should
be taken into account in order to decide what kind of critical
points actually exists. Apparently (in the replica symmetry
approximation) there is a critical point at
$(\varepsilon,\alpha)\simeq (0.3492,0.833)$ where the first order
line ceases to exist, and an endpoint nearby
(0.3543,0.7784) where both lines (first and second
order) cross (see fig.\ref{cp}). It must be emphasized that the
retrieval solution is unstable in this limit ($T=0$), as can be
seen by the negative value of the entropy
\begin{equation}
S_o \equiv \beta^2 \left. \frac{\partial f}{\partial \beta}
\right|_{T=0}  = - \frac{\alpha}{2} \ln [1-C(1-\varepsilon y)]
- \frac{\alpha}{2} \frac{C(1-\varepsilon y)}{1-C(1-\varepsilon y)}
\;\;\;\; .
\label{ent}
\end{equation}
The numerical values of the entropy are larger than in the
Hopfield model, possibly indicating that the effects of RSB here
are  stronger. There is a range of
$\varepsilon$, $0.3492 < \varepsilon < 0.3543$  where
the system suffers up to four phase transitions (two first and two
second order) since, for a fixed value of $\varepsilon$, as
$\alpha$ is increased it crosses twice the line
$\alpha_c'$. The richness of  this
phase diagram deserves a separate study including the
analysis of the stability of the solutions as well the effects of
RSB, mainly in this low temperature region, but it is beyond the
scope of this paper. The peak in the second retrieval region occurs
because the noise in the Hopfield term of the local field in
eq.(\ref{dynamic}) is completely compensated by the noise generated
by self-couplings in the fourth order term when
$\varepsilon=y^{-1}$ (see discussion below). For negative values
this never happens since $1+|\varepsilon| y$ is always positive
($y$ is positive defined). Differently from the GH model, here
there is no
cut-off in $\alpha_c'$ as $\varepsilon$ decreases, although
$\alpha_c'\to 0$.

The points where $m$ continuously approaches zero ($m \sim |\alpha
- \alpha_c^{\pm}(\varepsilon)|^{1/2}$ as $\alpha\to
\alpha_c^{\pm}(\varepsilon)$) ,
$\alpha_c^{\pm}(\varepsilon)$, are obtained by expanding
equations (\ref{mt0})--(\ref{ct0}) for small $m$:
\begin{equation}
\alpha_c^{\pm}(\varepsilon)=\left(\frac{1}{\sqrt{\varepsilon}} \pm
\sqrt{\frac{2}{\pi}}  \:\right)^{2}
\:\:\:\: .
\label{alfac}
\end{equation}
For $\varepsilon>\varepsilon_c$ there is only one transition at
$\alpha_c^+ (\varepsilon)$, as can be seen in the $T=0$ phase
diagram (fig.\ref{pdt0}), and the critical value of $\alpha$ is a
decreasing function of $\varepsilon$. Two different retrieval
phases appear for
$\varepsilon<\varepsilon_c$ and when the right
border is second order, given by $\alpha_c^-$, the width of the
second retrieval region is
\begin{equation}
\Delta' \equiv \alpha_c^+ (\varepsilon) - \alpha_c^-(\varepsilon)
=4 \sqrt{\frac{2}{\varepsilon
\pi}} \label{deltalinha}
\end{equation}
and the gap $\Delta$ between the first and second retrieval regions
is \begin{equation}
\Delta \equiv \alpha_c^-(\varepsilon) - \alpha_c'(\varepsilon)
\;\;\;\; ,
\end{equation}
where $\alpha_c'(\varepsilon)$ is the point of the first order
transition in the
first region. As $\varepsilon\to 0 $, $\Delta$ goes to
infinity as $\varepsilon^{-1}$ and $\Delta'$ as
$\varepsilon^{-1/2}$. Thus, when the Truncated model recovers the
Hopfield model ($\varepsilon\to 0$), the location of the second
retrieval region in the $\alpha$-axis goes to infinity and
$\alpha_c'\to 0.138$.

When $\varepsilon\to\infty$, the solution for $m$ can be obtained
as a function of $\alpha$ and it reads
\begin{equation}
m= \mbox{erf} \left( \sqrt{\frac{1}{2} \ln \frac{2}{\alpha\pi}}
\right)
\end{equation}
and
\begin{equation}
\alpha_c^+(\infty) = \frac{2}{\pi} \;\;\;\;\; .
\end{equation}
In this limit case, the system behaves suitably as an
associative device ($m\simeq 1$) only if the number of embedded
patterns is finite ($\alpha=0$).

It is also possible to obtain the location of the maximum value of
$m$ in the second retrieval region: the values of
$\alpha$ that allow $m=1$ in (\ref{mt0})-(\ref{ct0}) are
$\alpha=0$ and
\begin{equation}
\alpha = \frac{1-\varepsilon}{\varepsilon} \;\;\;\; .
\label{alfaopt}
\end{equation}
At the peak the noise from the high patterns (measured by $r$) goes
to zero because $\varepsilon=y^{-1}$ (and since $y$  is positive
defined, this only happens for positive $\varepsilon$). Thus, the
contribution from the second order term does not contribute in the
peak and one shall take into account the
next (fourth order) term.
This allows us to introduce an optimal learning rule in the
next section by choosing the weight $\varepsilon$ as the
value that satisfies (\ref{alfaopt}).

The above results can be qualitatively understood through a signal
to noise analysis. The local field acting
upon the $i$-th neuron when the system is recalling the
first pattern is
\begin{equation}
h_i \simeq (1-\alpha\varepsilon) \xi_i^1 + (1-\alpha\varepsilon -
\varepsilon) \frac{1}{N} \sum_{\mu\neq1} \sum_{j\neq i}
\xi_i^{\mu} \xi_j^{\mu} \xi_j^1 \;\;\;\; ,
\label{signalnoise}
\end{equation}
where we considered the contributions from both two and four neuron
couplings in eq.(\ref{liapunov}). The increase in $\varepsilon$ has
then two effects: it acts both on the signal and noise terms.
Depending on the value of $\varepsilon$, an increase in $\alpha$
may either suppress the signal or the noise term. When
$\varepsilon$ is small enough, the sum in eq.(\ref{signalnoise})
may overwhelm the signal term and we have a Hopfield-like mechanism
of
suppressing retrieval abilities. In this case the overall
behavior is qualitatively similar to the second order Hopfield
model, as in the GH model. This also corresponds to
the first retrieval region for $\varepsilon<\varepsilon_c$. On the
other hand, when $\varepsilon>\varepsilon_c$, together
with a decrease in the noise term, one can also observe a
detectable decrease in the signal term with increasing $\alpha$
when in the second retrieval region. Now the mechanism of loss of
retrieval ability is not due to an overwhelming noise term, but to
the annihilation of the signal one. In other words,
the minimum of energy is not shifted from the pattern due
to the noise, but the energy barriers around the minima
are decreased. A similar effect is observed in the complete model
\cite{rsana}. This explains the qualitatively diverse behavior of
the network in this region. Also, the second (noise) term  has zero
mean and variance
equal to $\sigma^2 = \alpha (1-\alpha\varepsilon -
\varepsilon )^2$. An important result is
that the variance is zero if $\alpha =0$ or
$\alpha=(1-\varepsilon)/\varepsilon$. These are
the points where the retrieval is perfect ($m=1$);
higher order noise terms, which we did not write, have
dispersions at least ${\cal O}(N^{-1})$ times smaller.
This will be considered in a next section when an
optimal learning rule will be detailed.
Also, as for increasing values of $\varepsilon$ the coefficient of
the signal term, $1-\alpha\varepsilon$, changes signal
for decreasing values of $\alpha$, explaining why the
critical value of $\alpha$ decreases.

To verify these results we performed zero temperature
simulations with network sizes up to 512 neurons.
The steps are the following: one of the embedded memories is chosen
as the initial state and a spin is (sequentially) flipped whenever
this  lowers the system energy. This procedure is
repeated until a stable fix point is reached and the final
overlap $m$ with the chosen memory is measured. The averages were
taken over 5 different sets of patterns and the number of runs in
each set were 200 and the sizes were $N=128,256$ and 512 neurons.
In fig.\ref{trssim} we
can see the final overlap $m$ versus $\alpha$ for
$\varepsilon=0.3$, clearly showing the existence of the gap.  The
results for $\varepsilon=1$ can be found in ref.\cite{trs}, but a
more extensive study of other values of $\varepsilon$ as well
dynamics differences between the first and second
regions is in course. As in other models,
when compared with the mean field calculations, the simulation
yields  some discrepancies (mainly near the transitions),
what is in part due to the replica symmetry instability at $T=0$
and supported by the negative entropy at zero
temperature obtained with the replica symmetry ansatz. Also remark
that the remanent magnetization above $\alpha_c^+$ for
$\varepsilon=0.3$,
$m\sim 0.1$, is lower than the one found for the Hopfield
and GH models ($m\sim 0.2$) and, as can be seen in fig.\ref{trsm2},
is a decreasing function of $\varepsilon$.

A final remark concerning the free energy of the
Truncated model at $T=0$: the
retrieval states are global minima for all values of
$\alpha$ below $\alpha_c^+$ in the Truncated model if
$\varepsilon>\varepsilon_c$. For $\varepsilon<\varepsilon_c$ there
is a range of $\alpha$ in the first retrieval region where they are
local  minima (metastable):
$\alpha_M < \alpha < \alpha_c'$ (in the second retrieval
region they are always global minima). For $\varepsilon\to 0$,
$\alpha_M\to 0.05$, as expected \cite{amit}.

On the other hand, in the GH model whose equations at $T=0$  are
(the equations for $m$ and $C$ are the same as (\ref{mt0}) and
(\ref{ct0}), respectively)
\begin{eqnarray}
t&=& \frac{1}{2} \sum_{\ell=2}^M \varepsilon_{\ell} \ell m^{\ell -
1} \label{thop} \\* r &=& (1-C)^{-2} \;\;\;\; ,
\end{eqnarray}
the storage capacity is a monotonically
crescent function of both $M$ and $\varepsilon_{\ell}$. Fig.
\ref{gh} shows $\alpha_c$ versus $\varepsilon$ in the case
$\varepsilon_{\ell}=\delta_{2\ell} + \varepsilon \delta_{4\ell}$,
that is, when only second and fourth order terms are considered.
The line  $\alpha_M$ where the
retrieval states become global minima and the value of $m$ at the
criticality, $m_c(\varepsilon)$, are also plotted.
When $\varepsilon\to\infty$, the asymptotic
value of $m_c(\varepsilon)$ goes to $m_c(\infty)=0.918$
and the critical value of $\alpha$ grows as
$\alpha_c \sim \varepsilon^2$ what can be
understood as follows. The bigger $\varepsilon$ is, the more
important the fourth order term and the system capacity tends to be
of order ${\cal O}(N^3)$ (attained in the
absence of the second order term \cite{multi}), that is, the
asymptotic behavior
of the maximum number of storable patterns goes as $P_c
\sim \varepsilon^2 N$. The value of $\alpha_M$ also
goes as $\varepsilon^2$ as $\varepsilon\to\infty$.
For negative values of $\varepsilon$ there is a cut-off:
below $\varepsilon^{cut}$ there is no retrieval ($t$, given  by
eq.(\ref{thop}), is null).
 For instance, for the case $\varepsilon_{\ell}=
\delta_{2\ell} + \varepsilon \delta_{k\ell}$, $\varepsilon^{cut} =
-2/k$. As $\varepsilon\to\varepsilon^{cut}$ (from above),
$\alpha_c\to 0$ and
$m_c\to 1$. In the general case we have
\begin{equation}
\sum_{\ell=2}^M \varepsilon_{\ell}^{cut} \ell = 0\;\;\;\; ,
\end{equation}
defining a hyper-plane in the space of the non-zero
$\varepsilon_{\ell}$'s.
There is also a cut-off in $\alpha_M$, below which
the memories are never
global minima of the free energy. For instance, for
the same case $k=4$, $\varepsilon^{cut}_M = 2/\pi -1 \simeq
-0.363$.

\subsection{The $T\neq0$ Case}

The GH model presents some universal features
that do not depend on the particular values of
$\varepsilon_{\ell}$ and $M$. For instance, the line below which
the SG states exist is
\begin{equation}
T_g = 1 + \sqrt{\alpha} \;\;\;\; , \;\;\;\; \forall
\varepsilon_{\ell}, M \;\;\;\;\;\;,
\label{tg}
\end{equation}
and $\varepsilon_2\neq 0$. The phase diagram for
$\varepsilon_{\ell}=\delta_{2\ell} + \delta_{4\ell}$ is
presented in fig.\ref{pdgh} ($\varepsilon=1$ since the
qualitative features do not change with $\varepsilon$).
There are three relevant lines: $T_g$, given by eq.(\ref{tg}),
signalling the appearance of SG states; $T_M$, where the retrieval
states ($m\neq 0$) first appear and finally $T_c$,
where the retrieval states become global minima of
the  free energy. At $\alpha=0$, $T_g$ and $T_M$ do not meet,
implying that there is a transition
between the retrieval and paramagnetic phase also for small values
of $\alpha$. This phase diagram is similar to that for
the $Q$-state Potts neural network model \cite{bolle},
due to a formal equivalence between multistates neurons and binary
spins with diluted multispin interactions \cite{kanter}.

The  SG phase is reentrant and hence the maximum possible value of
$\alpha$ is not at $T=0$ ($\alpha_c\simeq 1.556 $) but at a non
zero  value of  $T$: $\alpha_c^{max}\simeq 1.566$ for $T\simeq
0.126$. The degree of reentrance depends on $\varepsilon$: in the
Hopfield limit ($\varepsilon_{\ell}=\delta_{2\ell}$) it is very
small \cite{canning}.
As a consequence, a small amount of noise improves the
storage capacity of the system. This can also be observed in the
behavior of $m$ with $T$ near the reentrant region: the overlap
first increases before decreasing, indicating a small improvement
with thermal noise. These effects may be an
artifact of the  replica symmetry (supposed to be
stronger near the reentrant region): $\alpha_c^{max}$ is believed
to be a lower bound for the actual $T=0$ critical
capacity obtained when the replica symmetry is broken (the
reentrant  phase would then disappear). In other words,
$\alpha_c^{RSB} \geq \alpha_c^{max}$, what is supported by
numerical
simulations in the case $\varepsilon=1$ \cite{trs}.

The phase diagram for the Truncated model with the lines
$T_M$ and $T_c$ for $\varepsilon=1$ is shown in figure
\ref{pdt1}. In figure \ref{pdt} the line $T_M$ is shown for two
values of $\varepsilon$: 0.36 and 0.3. For $\varepsilon >
\varepsilon_c$, the retrieval states are always global
minima of the free energy for low temperatures, although
for $\varepsilon<\varepsilon_c$ they may become local minima
in the first retrieval region. For all values of
$\varepsilon$ and $\alpha=0$, $T_M=T_g=T_c=1$.
Nevertheless, for $\alpha\neq0$, the qualitative
features of the phase diagram depend on $\varepsilon$,
differently from the GH model. When $\varepsilon=1$
(fig. \ref{pdt1}), $T_M$ decreases monotonically and
there is no reentrant phase. As $\varepsilon$
decreases, the $T_M$ line develops a minimum, implying that for
some range of temperatures there are two retrieval
regions (e.g., for $\varepsilon=0.36$,
$0.086<T<0.202$), as shown in fig.\ref{pdt}.
For even smaller $\varepsilon$, the minimum becomes a gap
(fig.\ref{pdt}) and the first retrieval region is
reentrant. Also, the introduction of  thermal noise in the
system decreases the value of $\alpha$ at which $m$ has
a maximum in the second retrieval region (fig.\ref{mt}).
At $T=0$ there are three phase transitions lines in the
$\varepsilon\times\alpha$ plane: one first order,
$\alpha_c'$ and two second order ones $\alpha_c^{\pm}$ (fig.
\ref{pdt0}). For $T\neq 0$, all three phase transitions lines are
first order.

The temperature at which the SG solution ($m=0,q\neq 0$)
continuously disappear ($q\to 0$) is
\begin{equation}
T_g(\alpha,\varepsilon) = \frac{1+\sqrt{\alpha}}{\sqrt{\alpha}}
\left[ (1-\alpha\varepsilon)(1+\sqrt{\alpha}) -1 \right]
\;\;\;\;\; .
\label{tgtrs}
\end{equation}
These lines are shown in figure \ref{tgag} for several
values of $\varepsilon$. From the above equation we can see that
$T_g(0,\varepsilon)=1$ for all values of $\varepsilon$ and for
$\varepsilon=0$ one recovers the Hopfield line
$T_g(\alpha,0)=1+\sqrt{\alpha}$. The line $T_g$, for
$\varepsilon\neq 0$, goes to zero at (see inset of fig.\ref{tgag})
\begin{equation}
\alpha_g = \frac{1}{4} \left( 1 - \sqrt{1+\frac{4}{\varepsilon}}
\right)^2 \;\;\;\;\; .
\end{equation}
The same happens in the pseudo-inverse model \cite{ks}, although in
that case the transition is discontinuous and $\alpha_g
\simeq 0.363$. Although there is a solution with $q=0$ for the SG
phase along the line  $\alpha=\varepsilon^{-1}$, the critical line
$\alpha_g$, where the SG phase disappears, is such that $\alpha_g
\leq \varepsilon^{-1}$ (see fig.\ref{tgag}, inset).

We performed a numerical simulation to verify whether there is or
not a value of $\alpha$ above which the number of spurious states
suffers a sudden decreasing. Differently from the simulation
presented in the previous section, here the initial state is chosen
at random and, after the system reached a stable state, one
searches for the memory with the maximum overlap with that state.
The idea is that if the system starts from random positions, the
final state is either one of the
embedded memories or some spurious state, if any.
In order to quantify the results we define the following quantity:
\begin{equation}
{\cal M}  \equiv \: \left\langle \! \left\langle \: m_{m_o=1} -
m_{ris}   \: \right\rangle \! \right\rangle \: \;\;\;\;,
\end{equation}
where $m_{ris}$ is the mean final overlap when the initial  state
is random and $m_{m_o=1}$ is the mean final overlap
when the initial state is one of the memories. The symbol
$\: \left\langle \! \left\langle \: \;\; \: \right\rangle \!
\right\rangle \:$ stands for average over several sets of patterns
and initial states. Notice that
this quantity is proportional to the fractional ``occupation'' of
the phase space by the basins of spurious states.
Also, if $\alpha>\alpha_c$ both lines should merge since
there is no memory retrieval, that is, this quantity yields
relevant information only for $\alpha<\alpha_c$. The simulation
results for the Truncated model with $\varepsilon=0.5$  are
presented  in fig. \ref{mris}: ${\cal M}=0$
for $\alpha=0$ and for $\alpha$ roughly above 0.5. This result
supports the analytical prediction of the disappearance of the SG
states  above a given value of $\alpha$. However, the agreement is
only qualitative since the predicted value of $\alpha_g/\alpha_c$
for  $\varepsilon=0.5$, is 0.2. We expect this difference to be
an artifact of replica symmetry instability at low
temperatures. For sake of comparison, ${\cal M}$ is also
shown for the Hopfield and GH models, figs.\ref{mrishop}
and \ref{mrisgh}, respectively. In both cases, below
$\alpha_c$, ${\cal M}$ seems to increase both with
$\alpha$ and $N$. For the Hopfield model, fig. \ref{mrishop}, the
${\cal M}$ dependence on $\alpha$ is linear, the exponent depending
on $N$. For $\varepsilon=1$, ${\cal M}$ attains a minimum for small
$\alpha$ and grows for increasing values of both $\alpha$ and $N$.
Thus, except for the Truncated model, the
spurious states dominate the phase space landscape, either
by its increasing number or by increasing basins of attraction.

\section{THE OPTIMAL LEARNING RULE}

The second retrieval region in the Truncated model originates in
the competition between the noises of the Hopfield term
and the fourth order couplings. The value of $\alpha$ at which this
noise term is completely compensated is associated to a maximum in
the retrieval quality and depends on the value of $\varepsilon$.
When $\varepsilon$ is allowed to vary with $\alpha$, more
specifically, if
\begin{equation}
\varepsilon = \varepsilon_{opt}(\alpha)= \frac{1}{1 + \alpha}
\;\;\;\; , \label{epsopt}
\end{equation}
the system always works in the minimum noise region
(see eq.(\ref{signalnoise})):
the retrieval quality is maximum and the capacity of the network is
greatly enhanced. To estimate the limit in the load parameter of
such a model, we investigated higher order noise terms which are of
the order $\varepsilon\alpha / N^{3/2} \sim P N^{-3/2}/(P+N)$, that
is, in the thermodynamical limit, this term always goes to zero
regardless the asymptotic behavior of $P$. On the other hand, the
signal term is also modified by the fourth order couplings. The
thermodynamical limit for the signal term $1-\alpha\varepsilon$
when $\varepsilon= \varepsilon_{opt}$ and $P\sim N^k$ is given by
\begin{eqnarray}
\lim_{N\to\infty} (1-\alpha\varepsilon_{opt}) = \left\{
\begin{array}{ll}
1 & , \mbox{if  } k<1  \\
1/2 & ,\mbox{if  } k=1 \;\;\;\;, \\
0 & ,\mbox{if  } k>1
\end{array}
\right.
\end{eqnarray}
that is, $P\sim N^1$ (finite $\alpha$) to guarantee that the
signal term is not zero. In this optimal learning rule, the load
limit of the network originates in the annihilation of
the signal term and not in a decrease in the signal to
noise ratio.

A final remark about this optimal rule is that when $\alpha$
increases, providing it remains finite, $\varepsilon_{opt}$
decreases: the contribution from the fourth order term
decreases and the changes in the fourth order couplings, see
eq.(\ref{learning}), are smaller. In other words, the more the
network knows, the
easier it is to learn new patterns. Also, from the biological point
of view, the synapses of order higher than two are
corrections to the second order ones, what is indeed the
case here, since the role played for the fourth order
corrections depends on the value of $\varepsilon$ and
the network presents larger load capacities with decreasing
$\varepsilon$. The GH model storage capacity, on the other
hand, is an increasing function of the weight $\varepsilon$.

These results are valid for $T=0$, and the optimal
behavior is expected to hold at low temperatures. It
could be interesting to check what happens for higher
values of $T$.

\section{CONCLUSIONS}

We compared the effect of two different fourth order corrections to
the standard Hopfield model by considering two learning rules and
investigated their behavior calculating the
retrieval capabilities with and without thermal noise. The phase
diagrams for both models were presented, and
the strikingly different behaviors presented by them come
from the nature of the fourth order connections, that may or not
present mixed memory terms.

The original non-truncated model \cite{rsana,rssim} showed
an improved performance due to a strong reduction of spurious
states, with the consequent enhancing of the load capacity. The
limit of the storage capacity of the network originates in the
lowering of the energy barriers between memories and not in the
dislocation of the energy minima from the patterns; the retrieval
is then always perfect at $T=0$.
However, the order of the couplings (and their number)
increases with $P$
such that the ratio of information per synapse decreases.

Here, we introduced a truncated
model that shows a very rich behavior, summarized by the many phase
diagrams presented in previous sections. Besides several phase
transitions, either first or second order, the system may present
a gap separating two distinct retrieval phases. The first one
recovers the behavior of the standard Hopfield model in the
$\varepsilon\to 0$ limit.
On the other hand, in the second retrieval region an increase in
the load of the network acts lowering the signal term.
Thus, the limit in the capacity in this region is then due to
the effect of lowering the energy barriers, differently
from the Hopfield model. Possibly the SG
phase disappearing for $\alpha>\alpha_g$ is another consequence of
 suppressing the noise: there is no detectable
spurious states, what is supported by the numerical simulation.
When the gap is not present, there is
only one retrieval region that is more Hopfield-like
for small $\alpha$ but presents a $T=0$ second order transition
when losing its retrieval abilities. The second region,
when exists, presents a maximum in the curve $m$ versus $\alpha$
for $\varepsilon=(1+\alpha)^{-1}$. Hence, in general, when a
network is designed to work in a given range of storage
(provided $\alpha$ is finite), it is always possible to
choose some $\varepsilon$ that optimizes the retrieval.
In this case, similarly as the original model, there is no
second order noise
terms (for a convenient weight $\varepsilon$) and the load
capacity is limited by the annihilation of the signal term.  However,
other quantities, for instance, the size of the basins of
attraction and/or the convergence time, still deserve  further
investigations.

The effect of the sixth order term in the
expansion eq.(\ref{tobetruncated}) may also be considered. However,
we do not expect new effects to appear but for an increase in the
storage  capacity (or maybe more than one gap) and there may exist
at least one pair $(\varepsilon_4^{opt},\varepsilon_6^{opt})$ that
would improve the capacity of the model. Furthermore, the situation
when only  high order connections
are diluted is interesting from the biological point
of view, but the dynamics is hard to treat mathematically
\cite{tama,derrida}. A version of the model in which dilution is
present in both terms is being presently  studied.

The stability of the solutions together with
RSB effects should be studied,
mainly in the $T=0$ limit that surprisingly showed a very rich
phase diagram. An interesting problem is to investigate whether the
critical point and the endpoint merge or not in a tricritical one
when the replica symmetry is broken and then to obtain the critical
exponents near those points.

At last, we should point out that these results still have an
unclear biological relevancy, although they are interesting {\em
per se}. Maybe other fields that use spin Hamiltonians may be
benefitted by using this model. As an example, one can use
it as a fitness function in theoretical population genetics
\cite{kau}, where the gap might stand for some constraints that
cannot be satisfied by any species in that environment or for some
forbidden genetical traits.

\vskip 2\baselineskip
\noindent
{\large\bf Acknowledgments}: We acknowledge useful discussions with
 A.T. Ber\-nar\-des, J.R. Iglesias, N. Lemke, P.M.C. de Oliveira,
T.J.P.  Penna and F.A. Tamarit. Also, JJA thanks M.C.
Barbosa for discussions on critical points. Work
partially supported by brazilian agencies Conselho Nacional de
Desenvolvimento Cient\'{\i}fico e Tecnol\'ogico, Financiadora de
Estudos e Projetos  and Funda\c{c}\~ao
de Amparo \`a Pesquisa do Estado do Rio Grande do Sul.

\newpage
\noindent {\Large {\bf Figure Captions}}

\begin{figure}[h]
\caption{Overlap versus $\alpha$ for the Truncated model and
$\varepsilon=0.3,0.36$ and 0.5.  When
$\varepsilon\to 0^+$ then $\alpha_c'\to 0.138$ and
$\alpha_c^{\pm}\to\infty$, recovering the original Hopfield model.}
\label{trsm}
\end{figure}

\begin{figure}[h]
\caption{Overlap versus $\alpha$ for the Truncated model for
$\varepsilon=1,2,5$ and $\infty$. The retrieval quality decreases
for increasing values of $\varepsilon$ and as
$\varepsilon\to\infty$,  $\alpha_c\to 2/\pi$. The circles are the
results for the numerical simulation using $N=512$ (see text).}
\label{trsm2}
\end{figure}

\begin{figure}[h]
\caption{The $T=0$ phase diagram for the Truncated model. For
$\varepsilon>\varepsilon_c \simeq 0.3587$ the gap $\Delta=
\alpha_c^- - \alpha_c'$
disappears while for $\varepsilon\to 0$, the gap $\Delta\to\infty$
as $\varepsilon^{-1}$. The lines $\alpha_c^{\pm}(\varepsilon)$ are
second order while $\alpha_c'(\varepsilon)$ is first order.  The
dashed line ($\varepsilon_{opt}$) is the value of $\alpha\neq0$
where the peak ($m=1$) occurs. Inset: the values of
$\alpha$ at the first  order transition ($\alpha_c'$) for
$\varepsilon<0$.}
\label{pdt0}
\end{figure}

\begin{figure}[h]
\caption{Region near the point where the gap appears. The dashed is
a second order line while the solid one stands for
first order. The line $\alpha_c'$ ends in a critical point
at $\varepsilon\simeq 0.3492$ and both lines cross at an endpoint
at $\varepsilon\simeq 0.3543$. Inset: overlap
versus $\alpha$ near the critical point showing the second
first order transition for $\varepsilon=0.3493, 0.3492$ and 0.3491
(from left to right).}
\label{cp}
\end{figure}

\begin{figure}[h]
\caption{Overlap versus $\alpha$ for the Truncated model and
$\varepsilon=0.3$. The full curve is the $T=0$
solution of eqs.(\protect{\ref{mt0}})--(\protect{\ref{ct0}}) while
the points are obtained through numerical simulation (see text).
Notice the perfect retrieval at $\alpha=0$ and
$\alpha=(1-\varepsilon)/\varepsilon\simeq 2.33$.}
\label{trssim}
\end{figure}

\begin{figure}[h]
\caption{Critical values of $\alpha$ versus $\varepsilon$ at $T=0$
for the GH model. The dashed is the line below which the memories
are global minima of the free energy, $\alpha_M$. The overlap at
the  criticality, $m_c(\varepsilon)$, is shown  in the inset
and its asymptotic value, $m_c(\infty)$, is 0.918. There is a
cut-off for negative values of $\varepsilon$ (in this particular
case $\varepsilon^{cut}=-0.5$) where $\alpha_c\to 0$ and
$m_c\to 1$ as $\varepsilon\to\varepsilon^{cut}$ from above. There
is another cut-off below which the memories are never global minima
of the free energy: $\varepsilon^{cut}_M \simeq - 0.363$.}
\label{gh}
\end{figure}

\begin{figure}[h]
\caption{Phase diagram for the $\varepsilon=1$ GH model. Remark the
strong reentrant behavior for both $T_c$ and $T_M$.
 The line $T_g$ is the same
for all values of $\varepsilon$ and $M$. Notice that in this case
there is a transition between the retrieval phase and the
paramagnetic one for small $\alpha$.}
\label{pdgh}
\end{figure}

\begin{figure}[h]
\caption{Phase diagram for the Truncated model with
$\varepsilon=1$. Below the line $T_M$ we have $m\neq 0$
solutions (retrieval states) and these states become
global minima of the free energy below $T_c$ (dashed).
Notice that
there is only one retrieval region and no reentrant phase.}
\label{pdt1}
\end{figure}

\begin{figure}[h]
\caption{The line $T_M$ for the Truncated model with
$\varepsilon=0.3$ and 0.36. In the later case, it already shows the
structure of two retrieval regions although they are still
connected.  There is a range of temperature in which we have two
retrieval  regions:  $0.086<T<0.202$. For $\varepsilon=0.3$ both
regions separate  and the first retrieval region is reentrant.}
\label{pdt}
\end{figure}

\begin{figure}[h]
\caption{Overlap $m$ versus $\alpha$ for $\varepsilon=0.36$ at two
different temperatures. The peak is shifted to the left by the
thermal noise and the points $\alpha_c^{\pm}$ that are continuous
at $T=0$ are discontinuous at $T\neq0$.}
\label{mt}
\end{figure}

\begin{figure}[h]
\caption{Transition temperature for the SG solutions, $T_g $, for
several values of $\varepsilon$ in the Truncated model. As
$\varepsilon\to 0$, one
recovers the Hopfield line  $ T_g = 1 + \protect{\sqrt{\alpha}}$.
In the inset we show the points $\alpha_g$ where $T_g=0$,
as well as the line $\varepsilon^{-1}$ (see text).}
\label{tgag}
\end{figure}

\begin{figure}[h]
\caption{Simulation results for ${\cal M}\equiv\: \left\langle \!
\left\langle \: m_{m_o=1}-
m_{ris}  \: \right\rangle \! \right\rangle \:$ versus $\alpha/\alpha_c$
for the Truncated model and
$\varepsilon=0.5$. The maximum increases with the size $N$ of  the
network. For $0.5<\alpha<\alpha_c\simeq 4.893$,
${\cal M}=0$ signals the low occupancy of the phase space by the
spurious states.}
\label{mris}
\end{figure}

\begin{figure}[h]
\caption{Simulation results for ${\cal M}\equiv\: \left\langle \!
\left\langle \: m_{m_o=1}-
m_{ris}  \: \right\rangle \! \right\rangle \:$ versus $\alpha/\alpha_c$
for the Hopfield model.
Notice the  linear behavior below $\alpha_c$.}
\label{mrishop}
\end{figure}

\begin{figure}[h]
\caption{Simulation results for ${\cal M}\equiv\:
\left\langle \! \left\langle \: m_{m_o=1}-
m_{ris}  \: \right\rangle \! \right\rangle \:$ versus $\alpha/\alpha_c$
for the GH model and
$\varepsilon=1$. As in original Hopfield model, the phase space
occupation by the spurious states grows with both $\alpha$ and $N$,
although for small values of $\alpha$ it seems to have a minimum.}
\label{mrisgh}
\end{figure}

\end{document}